\documentclass[aps,prd,preprint,preprintnumbers]{revtex4}
\usepackage{amssymb}
\usepackage{amsmath}
\usepackage{verbatim}
\usepackage{mathrsfs}
\usepackage{amsfonts}
\usepackage{latexsym}
\usepackage{epsfig}
\usepackage{color}
\usepackage{cancel}
\usepackage{graphicx,subfigure}
\usepackage{units}
\begin{document}

\title{Scattering processes could distinguish Majorana from Dirac neutrinos}
%en orden alfabetico...
\author{J. Barranco}
\email[]{jbarranc@fisica.ugto.mx}
%\affiliation{Departamento de F\'isica, Divisi\'on de Ciencias e Ingenier\'ia, Campus Le\'on,
%Universidad de Guanajuato, Le\'on 37150, M\'exico}

\author{D. Delepine}
\email[]{delepine@fisica.ugto.mx}
%\affiliation{Departamento de F\'isica, Divisi\'on de Ciencias e Ingenier\'ia, Campus Le\'on,
%Universidad de Guanajuato, Le\'on 37150, M\'exico}
\author{V. Gonzalez Macias}
\email[]{vanniagm@fisica.ugto.mx}
%\affiliation{Departamento de F\'isica, Divisi\'on de Ciencias e Ingenier\'ia, Campus Le\'on,
%Universidad de Guanajuato, Le\'on 37150, M\'exico}
\author{C. Lujan-Peschard}
\email[]{carolup@fisica.ugto.mx}
\author{M. Napsuciale}
\email[]{mauro@fisica.ugto.mx}
\affiliation{Departamento de F\'isica, Divisi\'on de Ciencias e Ingenier\'ia, Campus Le\'on,
Universidad de Guanajuato, Le\'on 37150, M\'exico}

\begin{abstract}
It is well known that Majorana neutrinos have a pure axial neutral current
interaction while Dirac neutrinos have the standard vector-axial interaction. 
In spite of this crucial difference, usually Dirac neutrino processes differ from Majorana processes by a term 
proportional to the neutrino mass, resulting in almost unmeasurable observations of this difference. 
In the present work we show that once the neutrino polarization 
evolution is considered, there are clear differences between Dirac and Majorana
scattering on electrons. The change of polarization can be achieved in 
astrophysical environments with strong magnetic fields. 
Furthermore, we show that in the case of unpolarized
neutrino scattering onto polarized electrons, this difference can be relevant 
even for large values of the neutrino energy. 
\end{abstract}

\maketitle
There are still many open questions in particle physics, and
many of these involve the leptonic sector: what is the neutrino mass
scale and the neutrino mass hierarchy? why is the neutrino mixing 
matrix so different from quarks? is there CP violation in the neutrino
sector? Is the neutrino a Dirac or a Majorana particle? 
Pontecorvo called this last question {\it the central problem in neutrino
physics}.  
If neutrino is a Majorana particle, then the neutrino is 
identical to its own anti-particle. If this is the case, the neutrinoless 
double beta decay is possible \cite{Furry:1939qr}. If such a process is 
experimentally observed, it will be an undoubted signal of
the Majorana nature of the neutrino \cite{Schechter:1981bd}.
If on the other hand, the neutrino is a Dirac particle, then the antineutrino is a different particle than the neutrino. \\
There is another crucial difference between Dirac and Majorana neutrinos 
\cite{Kayser:1981nw,Mohapatra:1998rq,Campagne:1995np}.
If we consider the neutrino-electron scattering, either Dirac or Majorana,
the effective Lagrangian at low energies can be written as:
\begin{equation}
\mathcal{L}_{\nu e}=\frac{G_F}{\sqrt{2}}[\bar u_{\nu_\ell}\gamma^\mu(1-\gamma^5) u_{\nu_\ell}]
[\bar u_e \gamma_\mu(g_V^\ell-g_A^\ell\gamma^5)u_e]\,,
\end{equation}
where the coupling constants are given by
\begin{equation}
g_V^\ell=-\frac{1}{2}+2\sin^2\theta_W+\delta_{\ell e}\,,\quad g_A^\ell=-\frac{1}{2}+\delta_{le}\,,\quad \ell=e,\mu,\tau\,.\label{coupling}
\end{equation}
The amplitude for the neutrino-electron scattering in the Dirac case is:
\begin{equation}
\mathcal{M}^D(\nu_\ell e\to\nu_\ell e)=-i\frac{G_F}{\sqrt{2}}[\bar u_e^f\gamma^\mu(g_V^\ell-g_A^\ell\gamma^5)u_e^i][
\bar u_{\nu_\ell}^f\gamma_\mu(1-\gamma^5)u_{\nu_\ell}^i]\,, \label{amp_Dirac}
\end{equation} 
while for the Majorana case, since the neutrino is its own antiparticle, 
the amplitude will be:
\begin{equation}
\mathcal{M}^M(\nu_\ell e\to\nu_\ell e)=-i\frac{G_F}{\sqrt{2}}\left[\bar u_e^f\gamma^\mu(g_V^\ell-g_A^\ell\gamma^5)u_e^i\right]\left[\bar u_{\nu l}^f\gamma_\mu(1-\gamma^5)u_{\nu_\ell}^i-\bar v_{\nu_\ell}^f\gamma_\mu(1-\gamma^5)v_{\nu l}^i\right]\,.
\end{equation} 
If the neutrino is a Majorana particle, then the following identity is valid:
\begin{equation}
\bar v_{\nu_\ell}^f \gamma_\mu(1-\gamma^5) v_{\nu_\ell}^i=\bar u_{\nu_\ell}^f \gamma_\mu(1-\gamma^5) u_{\nu_\ell}^i\,,
\end{equation}
hence, the amplitude for the Majorana case will be
\begin{equation}
\mathcal{M}^M(\nu_\ell e\to\nu_\ell e)=i\frac{2 G_F}{\sqrt{2}}\left[\bar u_e^f\gamma^\mu(g_V^\ell-g_A^\ell\gamma^5)u_e^i \right]
\left[\bar u_{\nu l}^f\gamma_\mu\gamma^5u_{\nu l}^i\right]\,. \label{amp_Majorana}
\end{equation} 
It is clear that eq. \ref{amp_Dirac} is very different from eq. \ref{amp_Majorana}. 
Nevertheless, the neutrino mass is extremely small ($m_\nu < 2$ eV \cite{Beringer:1900zz}) 
thus these are almost completely chiral states, that is, 
almost fully polarized particles due to
the lefthanded nature of the charged weak interaction. For this reason, 
an extra state preparation factor  $(1-\gamma_5)/2$ is usually added such that 
eqs. (\ref{amp_Dirac}) and \ref{amp_Majorana} become identical \cite{Kayser:1981nw}.
The additional preparation factor is true if the neutrino mass is zero, but
once the neutrino mass is incorporated, the neutrino is not completely polarized.
Indeed, for instance, for the pure leptonic decay of a pseudoscalar 
meson $P^+ \to \ell^+ + \nu_\ell$, the neutrino longitudinal polarization is a 
function that depends on the neutrino mass \cite{Barenboim:1996cu}:
\begin{equation}
P_{\mbox{long}}=\frac{(E-W)|\vec k|}{W E-|\vec k|^2}\,,\label{longitudinal}
\end{equation}
with $W$ and $E$ the energies of the charged lepton $\ell$ and the neutrino 
respectively. 
$\vec k$ is given by $m_\nu^2=m_P^2+m_\ell^2-2m_P\sqrt{m_\ell^2+|\vec k|^2}$, 
$m_P$ the mass of the pseudoscalar meson, $m_\ell$ the lepton mass and
$m_\nu$ the neutrino mass.
For a neutrino mass of 1 eV, the polarization differs from a 
completely left handed lepton in one part in a billion, thus, the 
prescription of adding a preparation factor seems reasonable. 
Nevertheless, since we are in the high precision test of the standard model era, 
it is reasonable to evaluate the differences between
the Dirac and Majorana neutrino-electron scattering cross sections
considering that neutrinos are highly polarized but not completely polarized.
A straightforward calculation of the neutrino-electron scattering 
$\nu_\ell(p_\nu,s_\nu)+e(p_e)\to\nu_\ell(p_{\nu}')+e(p_e')$, with the
incident neutrino polarization vector defined as $s_\nu=(0,s_\perp,0,s_{||})$
in the neutrino rest frame , gives for the Dirac case \cite{Kayser:1981nw}:
\begin{eqnarray}
&&\frac{d\sigma^D}{d\Omega} = \frac{G_F^2}{8 \pi^2 s}
((m_e^2(E_\nu-p^2\cos\theta)({g_A^\ell}^2-{g_V^\ell}^2)\nonumber \\
&+& (E_\nu E_e+p^2)({g_V^\ell}+{g_A^\ell})^2+(E_\nu E_e+p^2\cos\theta)^2({g_V^\ell}-{g_A^\ell})^2 \nonumber\\ 
&-&p[s^{1/2}(E_\nu E_e+p^2)s_{||}({g_V^\ell}+{g_A^\ell})^2+(E_\nu E_e+p^2\cos\theta)\nonumber \\
&\times&\left((E_e+E_\nu\cos\theta)s_{||}+m_\nu s_{\perp}\sin\theta\cos\phi]\right)({g_V^\ell}-{g_A^\ell})^2\nonumber\\
&+&m_e(E_\nu(1-\cos\theta)s_{||}-m_\nu|s_{\perp}|\sin\theta\cos\phi)({g_A^\ell}^2-{g_V^\ell}^2)))\,,\label{Dirac}
%\righ]\right}
\end{eqnarray}
while for the Majorana case it is given by \cite{Kayser:1981nw}:
\begin{eqnarray}
&&\frac{d\sigma^M}{d\Omega}=\frac{G_F^2}{4\pi^2s}(((E_\nu E_e+p^2)^2+(E_\nu E_e+p^2\cos\theta)^2\nonumber\\
&+&m_\nu(E_\nu^2-p^2\cos\theta))({g_V^\ell}^2+{g_A^\ell}^2)+m_e^2(E_\nu^2-p^2\cos\theta+2m_\nu^2)\nonumber\\
&\times&({g_A^\ell}^2-{g_V^\ell}^2)-2{g_V^\ell}{g_A^\ell} p(2E_\nu E_e+p^2(1+\cos\theta))\nonumber\\
&\times&(E_\nu s_{||}(1-\cos\theta)-m_\nu|s_\perp|\sin\theta\cos\phi))\,.\label{Majorana}
\end{eqnarray}
In both cases, the variables $E_e,E_\nu,\theta,\phi$ and $p$ refer to center of mass (CM) quantities and
$s$ the Mandelstam variable. 
\begin{figure}
\includegraphics[width=1.\textwidth]{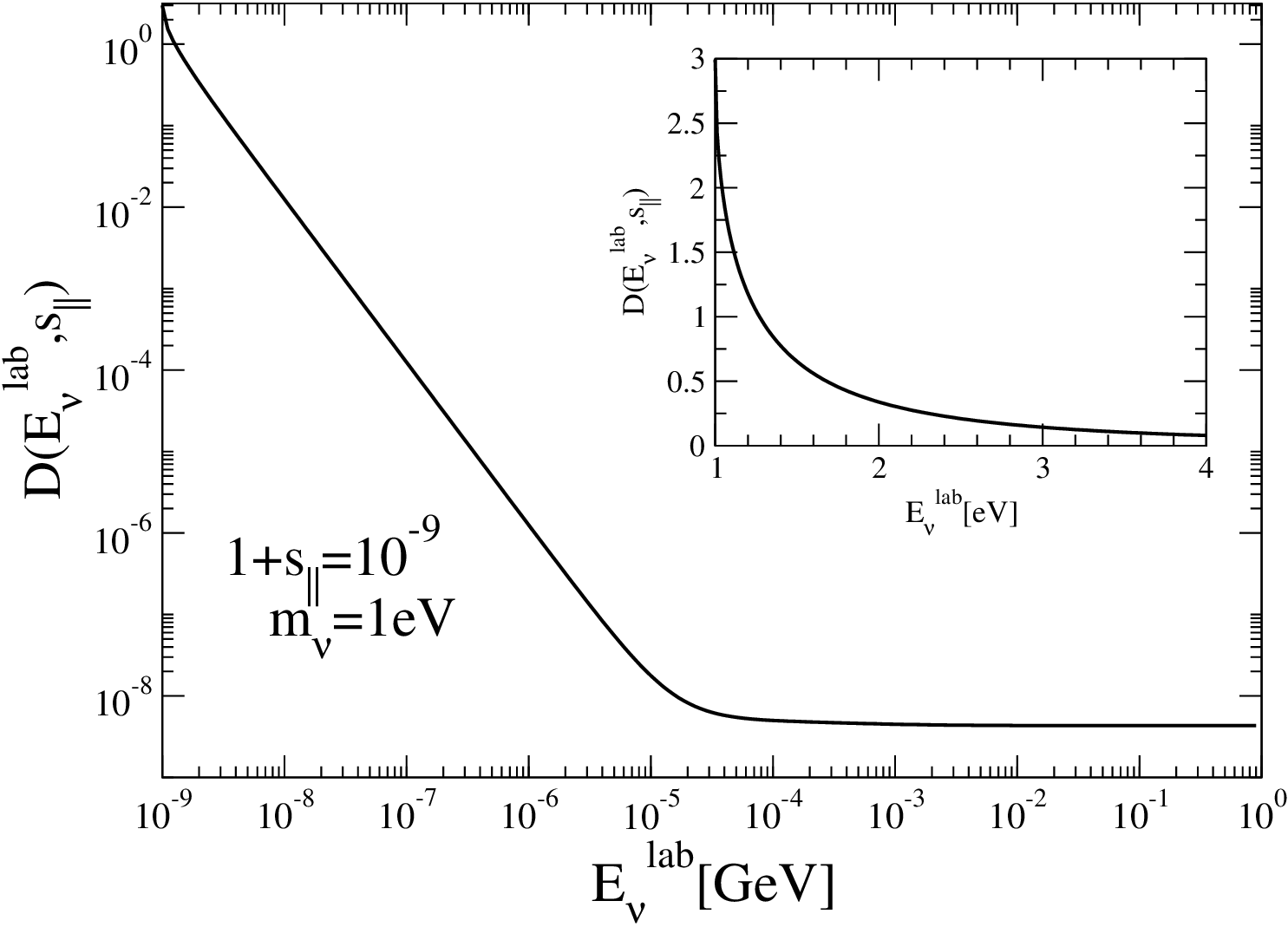}
\caption{Difference between the Majorana and Dirac neutrino-electron elastic scattering for
a longitudinal polarization $1+s_{||}=10^{-9}$ and a neutrino mass $m_\nu=1$ eV. Inner figure 
shows a zoom for low neutrino energies.}\label{Fig1}
\end{figure}
In order to quantify any difference between the Majorana and the Dirac cases, we define the function
\begin{equation}
D(E_\nu^{\mbox{lab}},s_{||})=\frac{\vert \sigma(\nu^D_{pol}e)- \sigma(\nu^M_{pol}e)\vert }{\sigma(\nu^D_{pol}e)}\,,\label{diferencia}
\end{equation} 
where we have integrated eqs. (\ref{Dirac}) and (\ref{Majorana}) over the CM angles and we have changed 
from the CM frame to the laboratory frame. 
This difference is shown in Fig. \ref{Fig1} and it summarizes a long discussion about the possibility
of distinguishing Dirac from Majorana neutrinos in neutrino-electron scattering processes 
\cite{Kayser:1981nw,Dass:1984qc,Garavaglia:1983wh,
Hannestad:1997mi}: {\it for terrestrial experiments where neutrinos are produced via charged currents, it is extremely difficult 
to observe significant differences between Dirac and Majorana neutrinos.} Indeed, as can be seen in Fig. \ref{Fig1},
for detectable neutrino energies, the difference is negligible. It becomes significant only for unreasonable
(of the order of eV) energies of the neutrino.  

Despite this fact, it is important to remember that neutrinos can forget its chiral origin. Indeed, 
any particle possessing a magnetic moment, as the neutrino does, interacts with external electromagnetic fields
and consequently, its spin may rotate around the direction imposed by this external field.
Furthermore, neutrinos can have a non negligible magnetic moment $\mu_\nu$. 
Actually, current experimental constraints only gives a superior bound on the neutrino magnetic moment,
$\mu_\nu < 3.2 \times 10^{-11}\mu_B$ \cite{Beringer:1900zz} with $\mu_B$ the Bohr magneton.
This limit is very big as compared with the expected neutrino magnetic moment that can arise from radiative corrections
in the Standard Model. For example, for a Dirac neutrino $\mu_\nu \sim 3 \times 10^{-19}\left(\frac{m_\nu}{1\mbox{eV}}\right)\mu_B$
\cite{Vogel:1989iv}. Hence, the spin of the neutrino could have a precession.
   
In the past, this spin precession has been used as a mechanism to probe the Majorana nature of the neutrino
\cite{Semikoz:1996up,Pastor:1997pb}, but those works have focused on a complete transition from
a Majorana neutrino to an antineutrino due to the spin precession. The non observation of
solar antineutrinos in KamLAND has derived an upper limit on the neutrino magnetic moment only $\mu_\nu<5\times 10^{-12}\mu_B$
\cite{Miranda:2003yh}.
\begin{figure}
\includegraphics[width=1.\textwidth]{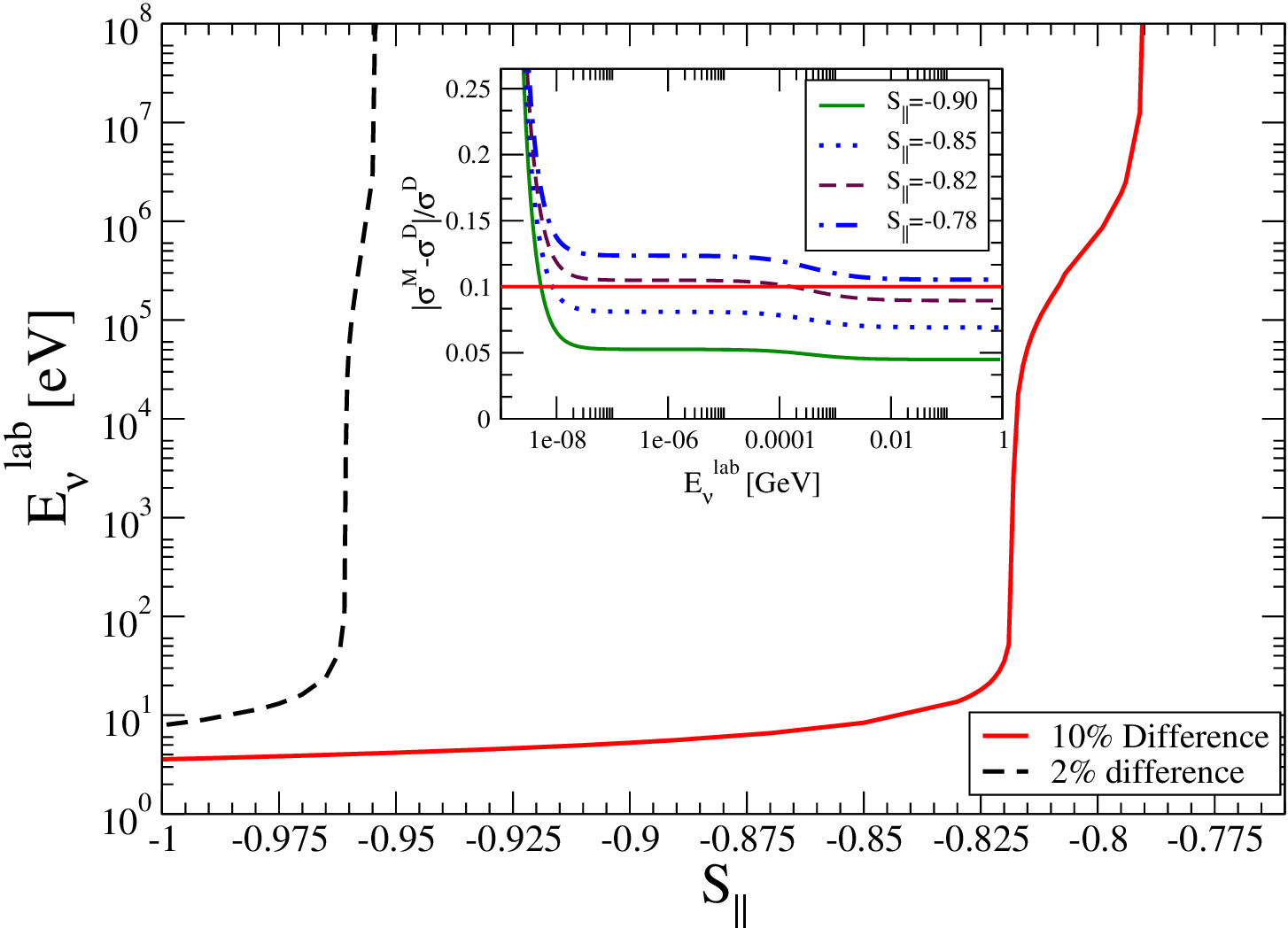}
\caption{Neutrino energy as a function of the neutrino longitudinal polarization needed
to have: $10\%$ difference between the Dirac and Majorana cross section (red solid line) and $2\%$
(black dashed line)}\label{Fig2}
\end{figure}

We will show, that contrary to those previous attempts, it is not necessary to have a completely flipped 
neutrino, that is an anti-neutrino,  to have observable differences between the Dirac and Majorana neutrinos. 
This can be observed by re-evaluating eq. \ref{diferencia} for different values of the neutrino longitudinal
polarization $s_{||}$. Inner plot of Fig. \ref{Fig2} shows the function $D(E_\nu^{\mbox{lab}},s_{||})$ for different
values of $s_{||}$.
As it can be seen, the value of the neutrino energy where a significant difference between Dirac and Majorana scattering
cross section appears is a function of $s_{||}$. In Fig. \ref{Fig2} we show the value of the neutrino energy for two cases:
when the difference is $10\%$ (solid red line) and $2\%$ (dashed black line). As it can be seen, there is an asymptotic value 
of $s_{||}$ where this difference is reachable for current neutrino detectors. In the first case, it is needed that the
neutrino forgets its chiral origin almost $25\%$, while a more precise experiment able to detect astrophysical 
neutrinos with a $2\%$ accuracy will need only a $5\%$ deviation in the neutrino's original helicity.

What is the magnetic field needed in order to have such changes in the neutrino's helicity?
In order to estimate this, we recall previous studies where the depolarization rate of neutrinos was calculated
\cite{Loeb:1989dr,Semikoz:1992yw,Elmfors:1997tt}. 
In the case of a random distribution of electromagnetic fields, the average neutrino's helicity $\langle h\rangle$
changes as dictated by the equation
%\begin{equation}
$\langle h(t)\rangle=exp(-\Gamma_{depol})\langle h(0)\rangle\,,$
%\end{equation} 
where 
\begin{equation}
\Gamma_{depol}=0.0132\mu_\nu^2 T^3\,, \label{random}
\end{equation}
$\mu_\nu$ the magnetic moment of the neutrino \cite{Elmfors:1997tt}.
Another source for neutrino spin depolarization is produced by the interaction with a large
scale magnetic field. For this case, the depolarization rate is given by
\begin{equation}
\Gamma_B=\frac{4\mu_\nu^2B^2 D}{\omega_{\rm refr}^2+4\mu_\nu^2B^2}\,,
\end{equation}
where $B$ is the external magnetic field, $\omega_{\rm refr} \simeq 1.1\times 10^{-20}(T/\mbox{MeV})^4$
for electron neutrinos, and $D \simeq 2.04 G_F^2 T^5$ \cite{Elmfors:1997tt}.

\begin{figure}
\includegraphics[width=1.0\textwidth]{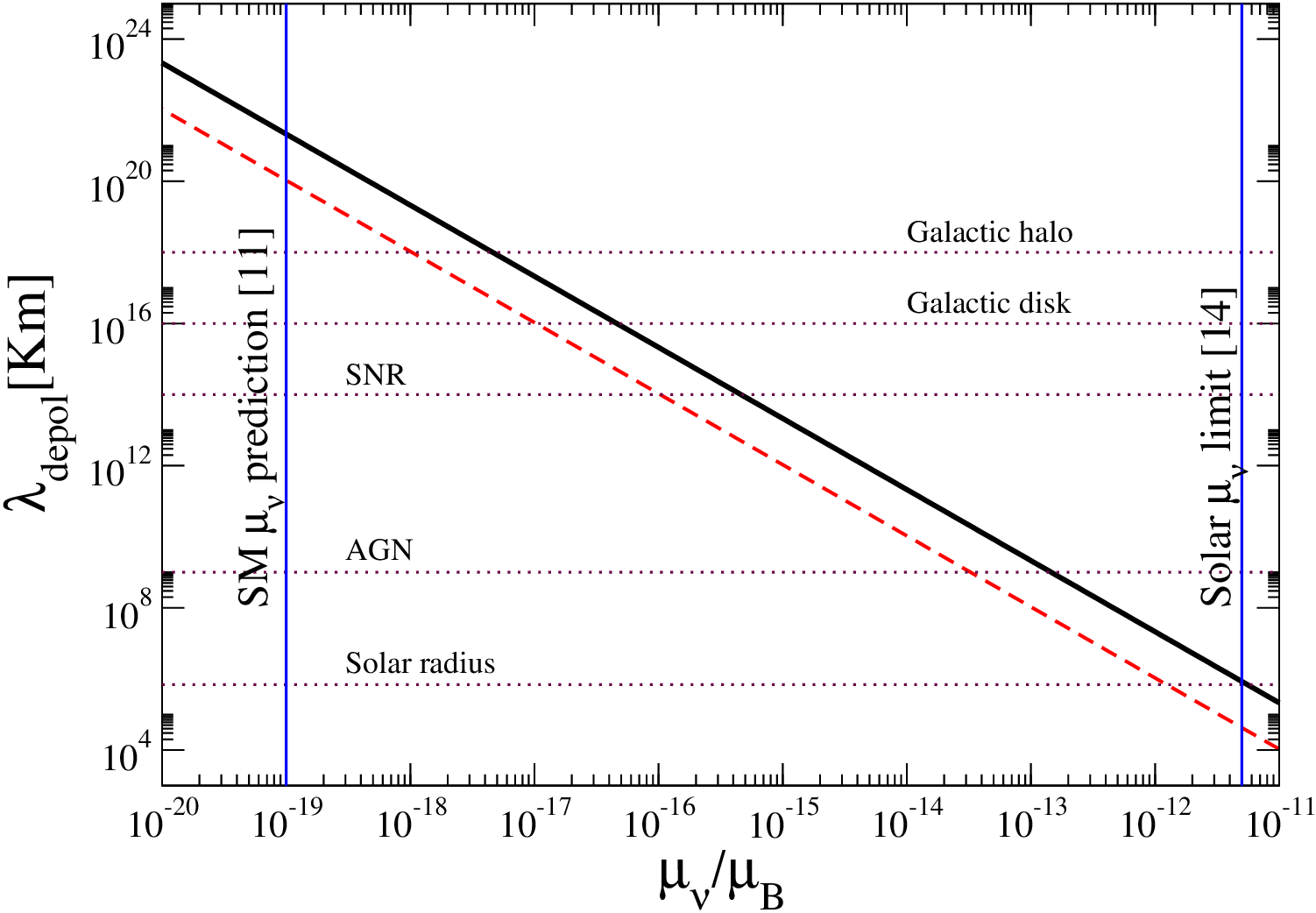}
\caption{Depolarization distance for a random distribution of magnetic fields as a function of the
neutrino magnetic moment. Solid black line is for a complete depolarization while red dashed line for a
$5\%$ change in the average neutrino helicity. Dotted lines represents typical astrophysical objects: The solar radius, 
the size of an Active Galactic Nuclei (AGN), a Super Nova Remant (SNR), the size of the galaxy disk and the galaxy halo.
We have assumed $T=20$MeV}\label{Fig3}
\end{figure}
Let us focus in the case of a random distribution of magnetic fields, eq. (\ref{random}).
In Fig. \ref{Fig3} we have plotted  the depolarization distance $\lambda_{depol}=1/\Gamma_{depol}$
as a function of the neutrino magnetic moment. Solid black line is for a complete depolarization while red dashed line for a
$5\%$ change in the average neutrino helicity. This is for a temperature of $T=20$ MeV.
Dotted lines represent typical astrophysical objects.
As expected, a complete neutrino - antineutrino oscillation due to the resonant spin precession
inside the sun gives the limit on the neutrino magnetic moment found in \cite{Miranda:2003yh}. 
This is represented by the intersection of the dotted line at the solar radius with the line 
of the neutrino depolarization. As it can bee seen, a more stringent limit could be obtained if
solar neutrino experiments reach a $2\%$ resolution in the antineutrino searches.

Moreover, Fig. \ref{Fig3} implies that other neutrinos produced in different astrophysical objects 
could in principle be depolarized if the neutrino magnetic moment is relatively large compared
to the SM prediction. 

Since neutrinos can have a broad distribution of spin polarization, finally, 
let us consider the extreme case where neutrinos are unpolarized and compute the 
neutrino-electron elastic scattering. As previously noted \cite{Hannestad:1997mi}, 
in this case the corresponding matrix elements for Dirac and Majorana neutrinos are
completely different. Furthermore, as eqs. \ref{amp_Dirac} and \ref{amp_Majorana} show,
{\it the difference increments as long as we maximize the axial contribution}. In order
to do that, we will consider that neutrinos are unpolarized and that the electron in 
the target is polarized.
The possible use of polarized electrons in $\nu-e$ elastic scattering has been previously 
proposed in order to look for other non standard neutrino interactions, as the 
neutrino magnetic moment itself \cite{Rashba:2000hv}, or possible $CP$ violation signals
\cite{Ciechanowicz:2003ds} among other different motivations \cite{Minkowski:2002sq}.
Here we will show that the electron polarization could enhance the difference between Dirac
and Majorana neutrinos as long as neutrinos had lost their initial polarization due to an
interaction of the neutrino with external magnetic fields in some wild astrophysical environment.

To take into account the electron polarization, we used the Michel-Wightman \cite{michel} formalism.
For the evaluations we used the laboratory frame, in which the electron is at rest 
and the electron polarization vector angle $\xi$ is given respect to the direction 
of the incoming neutrino (that we choose to be $z$): $s_e=(0,s_\perp,0,s_\parallel)$,  $s_{||}=\cos\xi$.

In the following we give the expressions for the differential cross section for the elastic scattering of $\nu_\ell-e$ 
considering the target electrons can be polarized in any direction. We will keep the dependence on the angle $s_{||}=\cos\xi$ 
with respect to the incoming neutrinos direction explicitly. 
\begin{figure}
\includegraphics[width=1.0\textwidth]{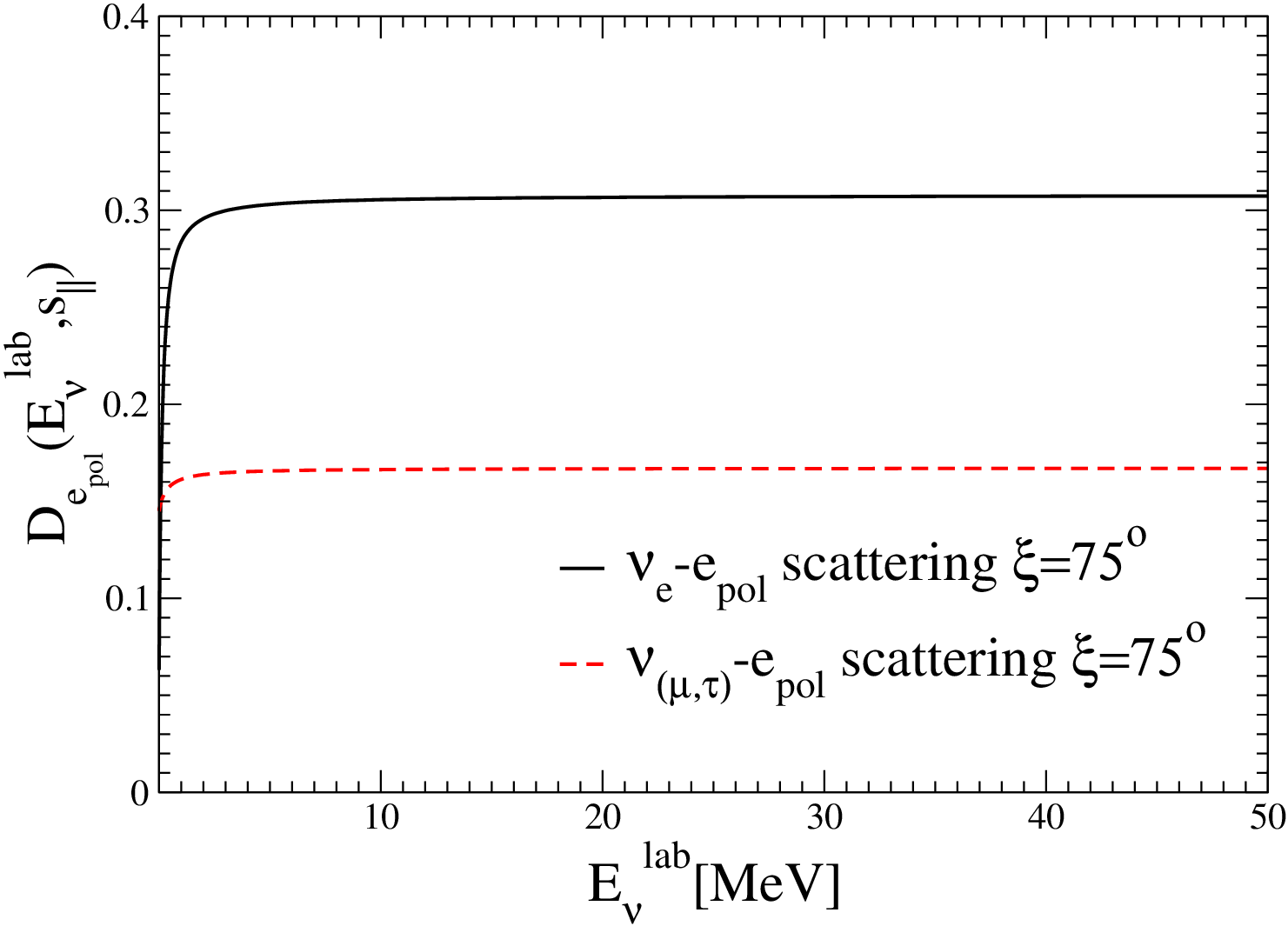}
\caption{Difference for the neutrino-polarized electron scattering for a fixed value of the
target electron polarization as defined by eq. \ref{diferencia2}. }\label{Fig4}
\end{figure}
We have neglected the neutrino mass, since the changes in the cross sections due to the inclusion of the neutrino mass
are very small.

We arrive to the Dirac neutrino-polarized electron elastic scattering:

\begin{eqnarray}{\label{dsdt}}
&&\frac{d\sigma(\nu_\ell^De_{pol})}{dT_e} =\frac{G_F^2 m_e}{2 \pi}  \left[  (g_A^{\ell} - g_V^{\ell})^2(1-s_{||}) 
\left(1 - \frac{T_e}{E_\nu^{lab}}\right)^2\right. \nonumber\\  
&+&\left.(1+s_{||})\left((g_A^{\ell}+g_V^{\ell})^2 +({g_A^\ell}^2 - {g_V^\ell}^2)\frac{m_e T_e}{{E_\nu^{lab}}^2}\right)\right.\nonumber\\ 
&+&\left.(g_A^{\ell}-g_V^{\ell})^2  s_{\vert \vert}  \left(1 - \frac{T_e}{E_\nu^{lab}}\right) \frac{m_e T_e}{E^2_\nu} \right]\,.      
\end{eqnarray}
Here, we have chosen the laboratory frame, with $T_e$ the electron recoil energy  as it was done in \cite{Rashba:2000hv},
and our results match. Recall $g_A^{\ell}\,, g_V^{\ell}$ are defined in eq. \ref{coupling}.
On the other hand, for the Majorana case, the neutrino-polarized electron  cross sections is given by:
\begin{eqnarray}
&&\frac{d\sigma(\nu_\ell^Me_{pol})}{dT_e}=\frac{G_F^2 m_e}{2 \pi}  \left[  2({g_A^\ell}^2 + {g_V^\ell}^2) \left(1 - \frac{T_e}{E_\nu^{lab}}\right)^2 
\right.\\ \nonumber
&+&\left.2 ({g_A^\ell}^2+{g_V^\ell}^2) + 2({g_A^\ell}^2 - {g_V^\ell}^2) \frac{m_e T_e}{{E_\nu^{lab}}^2} \right.\\ \nonumber
&+&\left.4 g_A^{\ell}  g_V^{\ell}  s_{\vert \vert}\left(1+\left(1-\frac{T_e}{E_\nu^{lab}}\right)^2-
\left(1-\frac{T_e}{E_\nu^{lab}}\right)\frac{m_eT_e}{{E_\nu^{lab}}^2}\right)  \right]
\end{eqnarray}

After an integration over the electron recoil energy, $0<T_e<\frac{2 {E_\nu^{lab}}^2}{m_e+2 E_\nu^{lab}}$ we can compute the
difference:

\begin{equation} 
D_{e_{\rm pol}}(E_\nu^{lab},s_{||})=\frac{\vert \sigma(\nu^{D}e_{pol})-\sigma(\nu^{M}e_{pol})\vert}{\sigma(\nu^{D}e_{pol})}\,,\label{diferencia2}
\end{equation} 
for different values of the electron polarization. An explicit example is shown in Fig. \ref{Fig4}.

It is remarkable that an appreciable difference is obtained for detectable energies of the
incoming neutrino.

\begin{figure}
\includegraphics[width=1.0\textwidth]{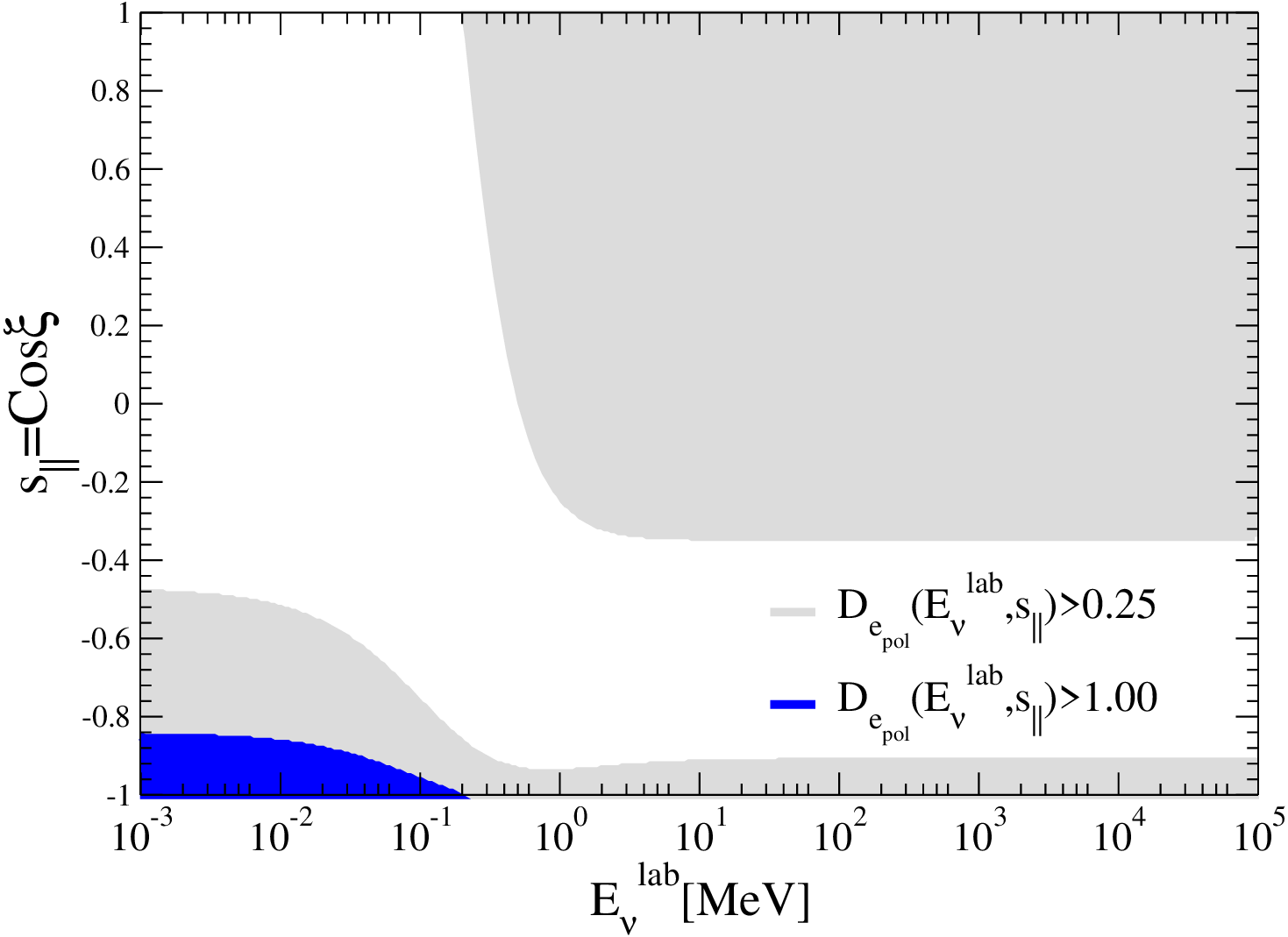}
\caption{Electron polarization needed to have significant differences between 
Dirac and Majorana cross sections for unpolarized neutrinos as a function
of the incoming neutrino energy.}\label{Fig5}
\end{figure}

Actually, in the case of unpolatized neutrinos that scatter on polarized electrons there could be
differences as big as twice the cross section, i.e. $D_{e_{\rm pol}}(E_\nu^{lab},s_{||})>1$, for certain values of the 
neutrino energy and the degree of polarization of the target electrons. This is illustrated in Fig. \ref{Fig5},
where we have plotted isocurves of $D_{e_{\rm pol}}(E_\nu^{lab},s_{||})$. Although this extreme case is reachable only for 
extremely low energetic neutrinos. Nevertheless, the case  $D_{e_{\rm pol}}(E_\nu^{lab},s_{||})>0.25$, i.e. differences of
at least $25\%$ are expected for a wide range in the electron polarization and neutrino energy.

In summary, we have shown that  once the neutrino polarization 
evolution is considered, there are clear differences between Dirac and Majorana
scattering on electrons. This change in the evolution of the helicity is
possible due to the existence of a neutrino magnetic moment. 
The change of polarization can be achieved in 
astrophysical environments with magnetic fields if the neutrino magnetic moment 
is bigger that the expected SM prediction but smaller than the current experimental 
limit. This helicity change could be strong enough to completely unpolarize the 
neutrino. Furthermore, we show that in the case of unpolarized
neutrino scattering onto polarized electrons, this difference can be relevant 
even for larger values of the neutrino energy.

\end{document}